\documentclass[fleqn,usenatbib]{mnras}

\usepackage{newtxtext,newtxmath}

\usepackage[T1]{fontenc}

\DeclareRobustCommand{\VAN}[3]{#2}
\let\VANthebibliography\thebibliography
\def\thebibliography{\DeclareRobustCommand{\VAN}[3]{##3}\VANthebibliography}

\usepackage{graphicx}	
\usepackage{amsmath}	
\usepackage{booktabs} 
\usepackage{xcolor}
\usepackage{hyperref}
\usepackage{xspace}
\usepackage{soul} 

\usepackage{glossaries}
\setacronymstyle{long-short}
\glsdisablehyper

\newcommand{\chron}{\mathcal{K}}
\newcommand{\Teff}{\,T_\mathrm{eff}}

\newcommand{\Msun}{\mbox{M$_{\odot}$}}
\newcommand{\Zin}{\mbox{$Z_{\rm in}$}} 
\newcommand{\dsct}{$\delta$\,Sct}

\newcommand{\NN}{neural network\xspace}

\usepackage{array}
\newcolumntype{H}{>{\setbox0=\hbox\bgroup}c<{\egroup}@{}}

\title[Asteroseismology of \dsct\ stars]{Asteroseismology of $\delta$ Scuti stars: emulating model grids using a neural network}

\author[Owen J. Scutt et al.]{Owen J. Scutt,$^{1}$\thanks{E-mail: oxs235@student.bham.ac.uk (OJS)}
Simon J. Murphy,$^{2}$\thanks{E-mail: simon.murphy@usq.edu.au (SJM)}
Martin B. Nielsen,$^{1}$
Guy R. Davies,$^{1}$
Timothy R. Bedding,$^{3}$
\newauthor
and Alexander J. Lyttle$^{1}$
\\
$^{1}$ School of Physics \& Astronomy, University of Birmingham, Edgbaston, Birmingham B15 2TT, United Kingdom\\
$^{2}$ Centre for Astrophysics, University of Southern Queensland, Toowoomba, QLD 4350, Australia\\
$^{3}$ Sydney Institute for Astronomy, School of Physics, University of Sydney NSW 2006, Australia
}

\date{Accepted XXX. Received YYY; in original form ZZZ}

\pubyear{2023}

\begin{document}
\label{firstpage}
\pagerange{\pageref{firstpage}--\pageref{lastpage}}
\maketitle

\begin{abstract}
Young $\delta$ Scuti stars have proven to be valuable asteroseismic targets but obtaining robust uncertainties on their inferred properties is challenging. 
We aim to quantify the random uncertainties in grid-based modelling of \dsct\ stars. 
We apply Bayesian inference using nested sampling and a \NN emulator of stellar models, testing our method on both simulated and real stars.
Based on results from simulated stars we demonstrate that our method can recover plausible posterior probability density estimates while accounting for both the random uncertainty from the observations and \NN emulation. We find that the posterior distributions of the fundamental parameters can be significantly non-Gaussian, multi-modal, and have strong covariance.  
We conclude that our method reliably estimates the random uncertainty in the modelling of \dsct\ stars and paves the way for the investigation and quantification of the systematic uncertainty. 

\end{abstract}


\section{Introduction}

Stellar ages for individual stars are notoriously difficult to measure \citep{soderblom2010}. One method is to model a cluster with isochrones, which is particularly sensitive to high-mass stars at the main-sequence (MS) turn-off \citep{lipatovetal2022}. Other techniques, such as the lithium depletion boundary (e.g. \citealt{galindo-guiletal2022}) or kinematics \citep{miret-roigetal2022,zerjaletal2023}, are able to use low-mass stars, which are much more abundant. However, methods that utilize intermediate-mass stars for measuring stellar ages have been lacking. 

Asteroseismology -- the study of stellar oscillations -- is highly sensitive to age and has long held promise as an independent method for age determination \citep[e.g.,][]{aerts2015}.  Like other techniques, asteroseismology is model-dependent, but the physics of those models is generally different from the high- and low-mass stars \citep{soderblom2010}, hence the techniques are highly complementary \citep{kerretal2022a,kerretal2022b}. Until recently, however, asteroseismology of intermediate-mass stars (the so-called $\delta$~Scuti variables) has been hampered by the difficulties in identifying which modes are excited. The discovery of regular patterns in the pulsation mode frequencies of some \dsct\ stars \citep{beddingetal2020} has opened up a pathway to determine their masses, ages, and metallicities, without the requirement that the star resides in a cluster or association. 

In recent years, oscillations in large numbers of \dsct\ stars have been measured using white-light photometry from space telescopes such as \textit{CoRoT} \citep[e.g.,][]{Paparo++2013, Michel2017, Barcelo-Forteza2018}, \textit{Kepler} \citep[e.g.,][]{uytterhoevenetal2011, Balona++2015, Garcia-Hernandez++2017, Bowman+Kurtz2018, Guzik2021} and \textit{TESS} \citep[e.g.,][]{antocietal2019, Hasanzadeh++2021, Barac++2022, Chen++2022}.
Observed oscillation frequencies can be compared against grids of model frequencies to find a best-fitting set of parameters \citep{murphyetal2021a,murphyetal2022a}. It is somewhat more challenging to understand the resulting uncertainties, which are not uniquely determined by the spacing of the model grid \citep{Pedersen2020}, and instead depend more strongly on the underlying physics \citep{steindletal2021b}. Part of the challenge is that models can be computationally expensive and calculating new evolutionary tracks on-the-fly for Monte Carlo sampling is prohibitive.

In order to treat the uncertainties more robustly, we aim to convert a discrete grid of stellar models into a continuous function. We use a \NN to emulate a grid of stellar models that has been pre-computed over the range of expected stellar parameters. We combine the trained \NN with a Bayesian sampler to formally treat random uncertainties in the observables. This yields estimates for the posterior probability density of the fundamental properties which quantifies their uncertainties. It also allows us to infer viable frequencies for modes that were not detected, but which might exist in the data at low signal-to-noise.

In the following section we describe the grid of stellar models on which the \NN is trained, and in Sec.\:\ref{sec:nn} we discuss the details of the network architecture and training method. In Sec.\:\ref{sec:inference} we present the method used to perform the Bayesian inference, and show results for a selection of simulated and real sets of observations (Sec.\:\ref{sec:results}). 
 
\section{The stellar model grid}
\label{sec:grid}
We used the model grid described in \citet{murphyetal2023}, consisting of evolutionary tracks computed with \textsc{MESA} (r15140; \citealt{paxtonetal2011,paxtonetal2013,paxtonetal2015,paxtonetal2018,paxtonetal2019}) and pulsation models calculated with \textsc{GYRE} (v6.0.1; \citealt{townsend&teitler2013}). Provisional versions of this grid have already been used to model the pulsations of \dsct\ stars \citep{murphyetal2022a,kerretal2022a,kerretal2022b,currieetal2023a}, and the physics of the models are described in \citet{murphyetal2022a}.

\begin{figure*}
    \centering
    \includegraphics[width=\linewidth]{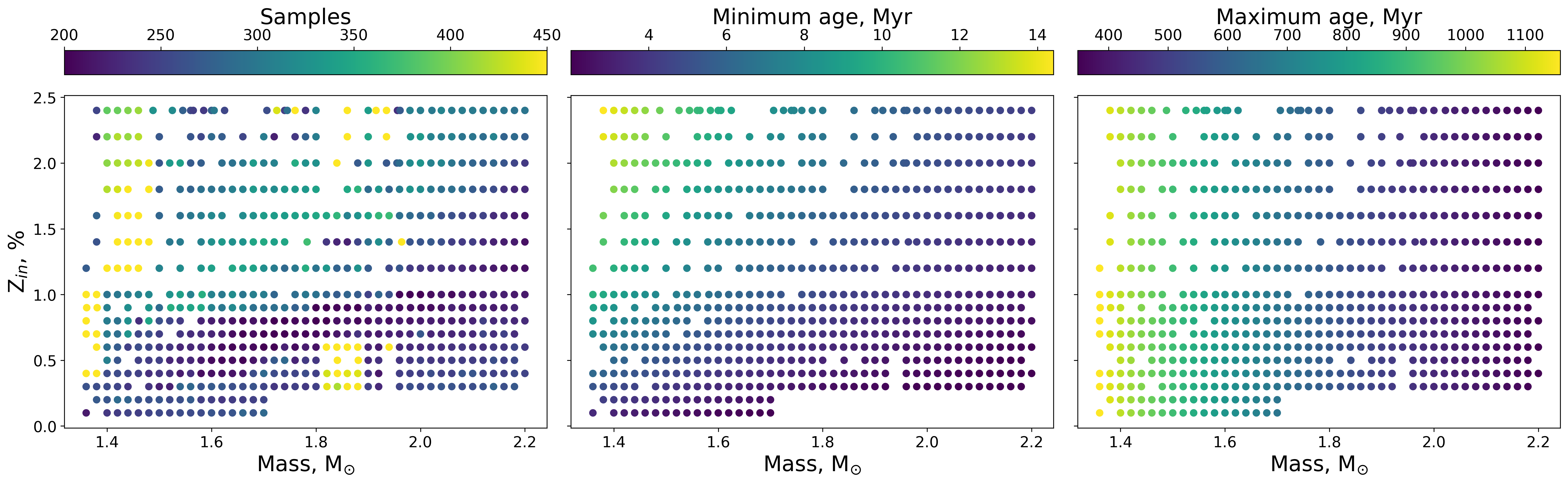}
    \caption{The model grid used in this work, where each symbol represents an evolutionary track with a particular metallicity ($Z_{\rm in}$) and mass. Colour-coding indicates the number of models along each track (left), and the minimum (middle) and maximum (right) ages for which pulsation frequencies were calculated (see Sec.~\ref{sec:grid}).}
    \label{fig:grid}
\end{figure*}

A well-sampled grid was needed to train the \NN emulator. Here, evolutionary tracks were spaced by 0.02\,\Msun\ in mass $M$ and 0.001 in initial metallicity \Zin. For $\Zin>0.010$, the spacing was increased to 0.002. The grid is shown in mass--metallicity space in Fig.\,\ref{fig:grid}. A common problem in \textsc{MESA} is that pre-MS models sometimes fail to converge and the evolution is terminated (see, e.g., \citealt{steindletal2021b}). In such cases, we attempted to re-calculate the track with a slightly increased mass ($M$+=\,0.001) up to five times before abandoning that track. Abandoned tracks appear as gaps in the grid in Fig.\,\ref{fig:grid}.

It is also important to ensure the tracks are sampled well in age. Computational errors are minimised by keeping the time interval small throughout the evolution, even if not all time steps are saved as outputs. The internal sampling is described in \citet{murphyetal2023}. For outputs, we saved evolutionary and pulsation models every 0.05\,Myr from 2\,Myr until 10.5\,Myr, in order to adequately sample the rapid evolutionary changes that occur on the pre-MS. After this the evolution is somewhat slower, and sampling of 3\,Myr was deemed adequate up to ages of 40\,Myr.  Beyond this, the tracks were instead sampled according to changes in position on the HR diagram (limits of $\Delta\log T_{\rm eff}=0.0006$ and $\Delta\log L=0.002$), with an upper limit of 100\,Myr between samples. The minimum and maximum ages sampled for each track are shown in Fig.\,\ref{fig:grid}. Where large gaps occurred in the grid, or when the specific $M$--\Zin\ combination demanded it, we manually recalculated tracks with finer sampling. This explains the variations in the number of samples per track in Fig.\,\ref{fig:grid}. The resulting grid spans classical observable ranges of $\log L=0.3$--1.8\,dex and $T_{\rm eff}=6000$--14000\,K.

For each pulsation model, we computed the frequencies of radial modes (spherical degree $\ell=0$) having radial orders $n$ from 1 to 11, and dipole ($\ell=1$) modes having $n\sim1$--10. This encompasses the range of radial orders observed for real stars \citep[e.g.][]{beddingetal2023}, with oscillation frequencies spanning $\nu=4$--150\,d$^{-1}$.
We calculated the mean frequency separation between radial orders ($\Delta\nu$) using the radial modes having $n=5$--9, by fitting a straight line to the mode frequencies as a function of $n$ \citep[see][]{White++2011}, using
\begin{eqnarray}
\nu = \Delta\nu(n+\ell/2+\epsilon).
\end{eqnarray}
The variable $\epsilon$ is the intercept of that line with the y-axis, and describes the distance of the radial mode ridge from the y-axis in an \'echelle diagram. In addition to the individual mode frequencies, we stored the values of $\Delta\nu$ and $\epsilon$ for each model in the grid, since these asteroseismic quantities relate to astrophysical quantities (\citet{murphyetal2023}).

To reduce the effect of the strong covariance between stellar age $\tau$ and mass $M$, and ease the training of the \NN, we used the assumption that the MS lifetime is approximately proportional to $M^{-3.2}$ and defined the scaled age \citep[e.g.,][]{2016AN....337..774D}
\begin{eqnarray}
\chron = 10^{-4}\, \tau\, (M/\Msun)^{3.2}.
\end{eqnarray}
This scaled age serves as an estimate of the fractional MS age of our models.

\section{Constructing the neural network}
\label{sec:nn}
To overcome the discretely sampled nature of the model grid, we used a \NN consisting of a series of fully connected dense layers in place of standard interpolation for continuous stellar model emulation. The network was trained on the model grid, learning to predict observable parameters given stellar model input parameters. This way, the network learned the map from model parameters to observables and could be used for likelihood estimation during inference. To this end, we used the fundamental parameters $M$, $Z_{\mathrm{in}}$, and scaled age ($\chron$) as inputs for parameter augmentation and network training. Outputs consist of the classical observables ($L$ and $T_{\rm eff}$); asteroseismic quantities ($\Delta\nu$ and $\epsilon$); and 11 radial and 10 dipole mode frequencies. We refer to these 25 outputs collectively as the `observable parameters'. 

Once the input and output parameters were defined, we carried out dataset-wide parameter augmentation to improve the training of the network. We converted all parameters (excluding $\epsilon$) to the decimal logarithm and applied a Z-score standardisation to all parameters (including $\epsilon$). Both of these operations restricted all parameters to similar ranges, to avoid the \NN assigning erroneously high importance to parameters spanning several orders of magnitude during training. We found the combination of the two operations to be optimal for this investigation. 

To further simplify the training process, we adopted a process of pre-training dimensionality reduction and in-network dimensionality reprojection similar to that described in \citet{Spurio_Mancini_2022}. We performed principal component analysis on the observable parameters in the model grid, as follows. For all models, we calculated the covariance matrix of all 25 observable parameters. The eigenvectors of the resulting covariance matrix, or `principal components', were ranked in order of descending eigenvalue, returning a list of principal components explaining the most to the least variance in the observable parameters. We determined how many principal components to include using the explained variance ratio, which describes the percentage of the variance of the observable space present in just the chosen principal components. We found that 9 principal components explained all but $10^{-4}$ per cent of the total variance. This sufficiently explained the covariance of the 25 parameters in the full observable space. 

The use of principal components presents the \NN with a simpler map to learn\textemdash replacing the fundamental parameters by the reduced dimensions of the `latent parameters'\textemdash and also removes covariance information from the observables, which is redundant for the \NN. We then added a custom non-trainable layer to the \NN, which projects the latent parameters back into the full observable parameter space before the network outputs predictions.

Finally, we split the model grid into a training and a testing set for the \NN. The training set was randomly selected to comprise 80 per cent of the model grid, to be seen by the network during training. The test set, composed of the remaining 20 per cent of the grid, was unseen by the network during the training process and was used solely for evaluation of network prediction performance. This served as a check that the network is capable of model grid interpolation --- the training set became a sparser representation of the original model grid, with the test set providing models guaranteed to hold combinations of parameters previously unknown to the network.

In addition to the data augmentation prior to training, the hyperparameters of a \NN can be tuned to promote faster and more stable learning. To quantify network performance for comparison between different hyperparameter permutations, we compared their validation loss profiles over multiple network training sessions. We adopted a `grid search' method for testing potential combinations of network hyperparameters. This involved creating a grid of potential values for the number of fully connected dense layers (ranging from 3 to 8 in steps of 1), activation functions, optimizers, learning rates, loss functions, and batch sizes. We populated a grid with these hyperparameters, and then tested the resulting network at each position in the hyperparameter grid for successful validation loss minimisation.

We found the optimal network consisted of 6 fully-connected dense layers of 64 neurons, each using an exponential linear unit activation function \citep{clevert2015fast}, followed by the custom layer for projection from latent to observable parameters, and a final dense output layer with linear activation function. We used the \textsc{Adam} optimizer \citep{kingmaetal2014} with a learning rate of $10^{-4}$, and the mean-squared-error loss function. For our training set of $1.4 \times 10^5$ models, a batch size of $6 \times 10^4$ models provided a good compromise between speed and training stability. We used a validation split of 25 per cent of the training set --- where the test set is used to evaluate \NN success after training, a `validation set' is used for evaluation of \NN success during training. Once primary training was complete with the learning rate above, we saved and recompiled the network with a slower but less volatile learning rate of $7 \times 10^{-5}$, and restarted training until no validation loss reduction was observed for $10^4$ training epochs. The network and custom latent-to-observable projection layer were constructed using the \textsc{TensorFlow} sequential API \citep{tensorflow2015-whitepaper}.

\begin{figure*}
   \centering
   \includegraphics[width=\linewidth]{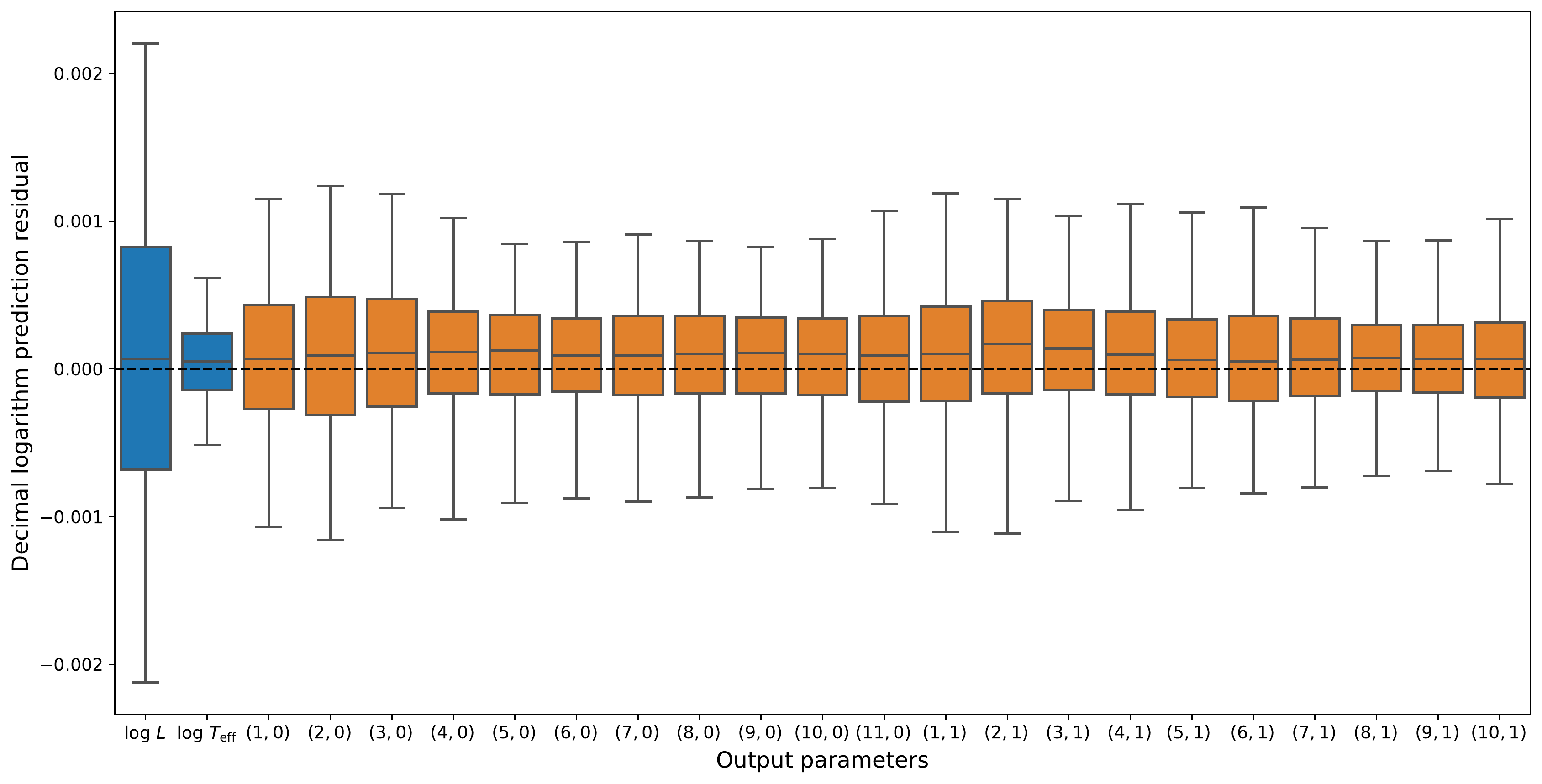}
   \caption{Residual distributions of predictions from the \NN.  The residuals are that of the decimal logarithms of $T_{\mathrm{eff}}$, $L$ (blue), and mode frequencies ($n,\ell$) (orange). The boxes show a central line at the median value of the distribution, with edges at the lower and upper quartiles. Whiskers extend to the 5th and 95th percentile range. The dashed line indicates complete agreement between the network predictions and model grid values.}
   \label{fig:nn_res}
\end{figure*}

\begin{figure}
    \centering
    \includegraphics[width=\linewidth]{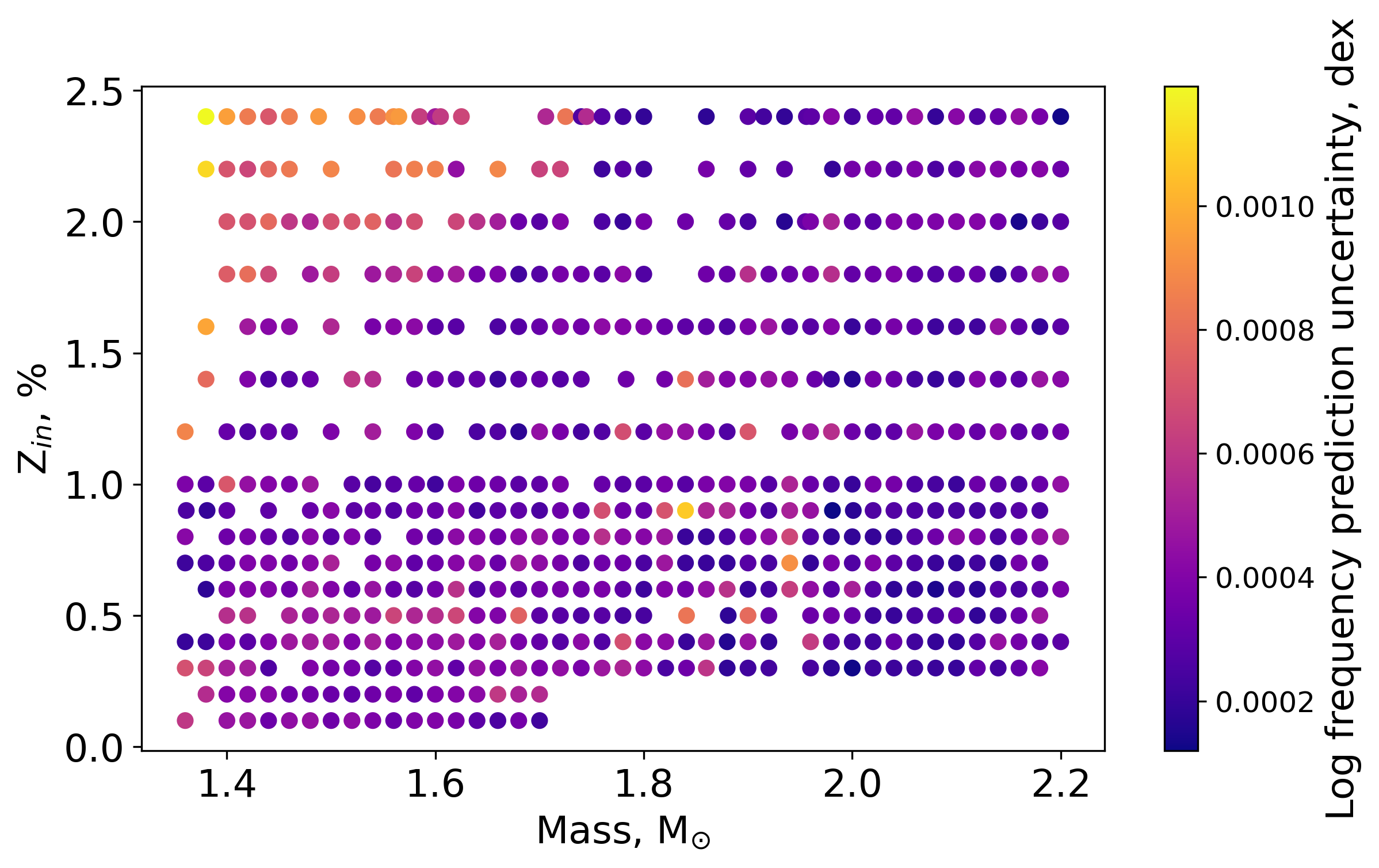}\
    \caption{Stellar model grid in model parameter space, coloured according to mean decimal logarithm of network uncertainty on mode frequency prediction averaged over each track.}
    \label{fig:unc_over_track}
\end{figure}

Once the optimal network from the grid search was trained, we evaluated the network performance across the full observable parameter space. Using the test set previously removed from the model grid, we plotted distributions of the decimal logarithm prediction residuals for each parameter. This allowed us to visualise any bias and uncertainty inherent in the \NN predictions. We used the median absolute deviation of these prediction residual distributions, shown in Fig.\,\ref{fig:nn_res}, to quantify network prediction uncertainty for observable parameters. We found an uncertainty in network predictions of $8\times10^{-4}$ dex and $2\times10^{-4}$ dex for $\log L$ and $\log T_{\mathrm{eff}}$, respectively. We found a mean prediction uncertainty of $\sim3\times10^{-4}$ dex for the decimal logarithm of the mode frequencies averaged over the entire test set. As shown in Fig.\,\ref{fig:unc_over_track}, \NN prediction uncertainty increased into the low mass, high metallicity domain. While the vast majority of the emulation uncertainty over tracks are well explained by our adopted uniform value, we note that \NN prediction uncertainty could be improved by better populating this region of the grid with more models for training.

\section{Inferring the fundamental stellar parameters}
\label{sec:inference}
To perform the Bayesian inference on the input model parameters, $\theta=\{M, \log Z_{\mathrm{in}}, \log \chron\}$, for a given observed set of mode frequencies, we sampled the posterior distribution
\begin{equation}
    P(\theta|D) = \frac{\mathcal{P}(\theta) \mathcal{L}(D|\theta) }{\mathcal{E}(D)}.
    \label{eq:posterior}
\end{equation}
Here, $\mathcal{P}(\theta)$ is the prior on the input model parameters, and $\mathcal{L}(D|\theta)$ is the likelihood of observing a set of parameters ($D$) given the model parameters. The evidence, $\mathcal{E}(D)$, is calculated at each step during the sampling. 

In addition to the input model parameters, we also included a variable offset term, $\Delta n$, as input to account for the possible ambiguity in assigning the radial orders of the observed modes. This ambiguity arises because the radial orders of a set of modes cannot always be determined from the observed mode frequencies alone, and are typically decided by comparison to stellar models. Including $\Delta n$ in the sampling allows us to marginalize over this uncertainty when estimating the posterior distribution of the input model parameters. We expect this uncertainty to only lead to an error of $\pm 1$ radial order, and so we chose the prior on $\Delta n$ to be a set of $\delta$-functions at $\Delta n = -1, 0 \, \mathrm{and}\,1$.

Table \ref{tab:priors} lists a summary of the prior functions. For the priors on $M$, $\log Z_{\mathrm{in}}$ and $\log \chron$, we chose to use $\beta$ distributions, since they can be bounded to match the limits of the stellar model grid. In addition, the shape parameters of the $\beta$ distributions may be chosen such that the priors reflect our expectation of the distribution of real observations of $M$, $Z_{\mathrm{in}}$ and $\chron$. 

Our choice of range and shape of the prior on $\chron$ is motivated by the age range and distribution we expect to target, and also be able to observe. Mode identification for \dsct\ stars is currently possible up to approximately one third of the MS age, after which the coupling between the buoyancy dominated and acoustic modes spoils the regular mode frequency patterns. Furthermore, at older ages the physics of mixing and overshooting become more important, and those were not treated as variables in the model grid. Hence, the prior on age extends to approximately one third of the expected MS lifetime. The lower limit on the prior on the scaled age was chosen because stars in our mass range of interest (see below) do not evolve to cross the \dsct\ instability strip until ages $\geq2$\,Myr. Due to the motion of stars through the \dsct\ instability strip, the age prior is biased toward lower ages, with a fall-off in the age prior distribution toward older stars. We note that a minority of points defined by the prior on $\log \chron$ do fall outside of the range of the grid. We define a $\beta$ distribution as our prior on $\log \chron$ with upper and lower limits of $-3$ and $-0.3$, respectively, whereas the model grid has a range of $-3$ to $-0.36$. This is not of concern as this high region of $\log \chron$ is extremely underpopulated - for example, the range  $-0.4 < \log \chron < -0.3$ contains just $0.4$ per cent of points in the entire grid. Furthermore, the \NN is capable of emulating points that lie outside the ranges of the training grid. In this regime, interpolation becomes extrapolation, and predictions are made with increased uncertainty. The prior region that lies outside of the grid is rapidly diminished during inference, and thus should not have bearing on the resulting model parameter posteriors.

The prior on \Zin\ ranges from approximately $0.07$ to $1.5$ times the solar metal mass fraction of $1.42$ per cent used in the models \citep{asplundetal2009}, which covers the metallicity distribution of stars forming within approximately 1\,kpc of the Sun at the current age of the Galaxy \citep{haydenetal2020}. The existence of metal-poor \dsct\ stars in modern star-forming regions (e.g. HD\,139614 in Upper Centaurus Lupus; \citealt{murphyetal2021a}) suggests that slightly sub-solar metallicities are more common than slightly super-solar metallicities in young \dsct\ stars. We therefore skewed the prior probability density toward sub-solar values.

Finally, the mass range was chosen to ensure that models exist both within and on either side of the instability strip \citep{dupretetal2004,murphyetal2019}. Our slight skew towards lower masses accounts for the similar skew present in the stellar initial mass function \citep{krumholz2014}. 

\begin{figure*}
    \centering
    \includegraphics[width=0.535\linewidth, trim={12 0 20 0}, clip]{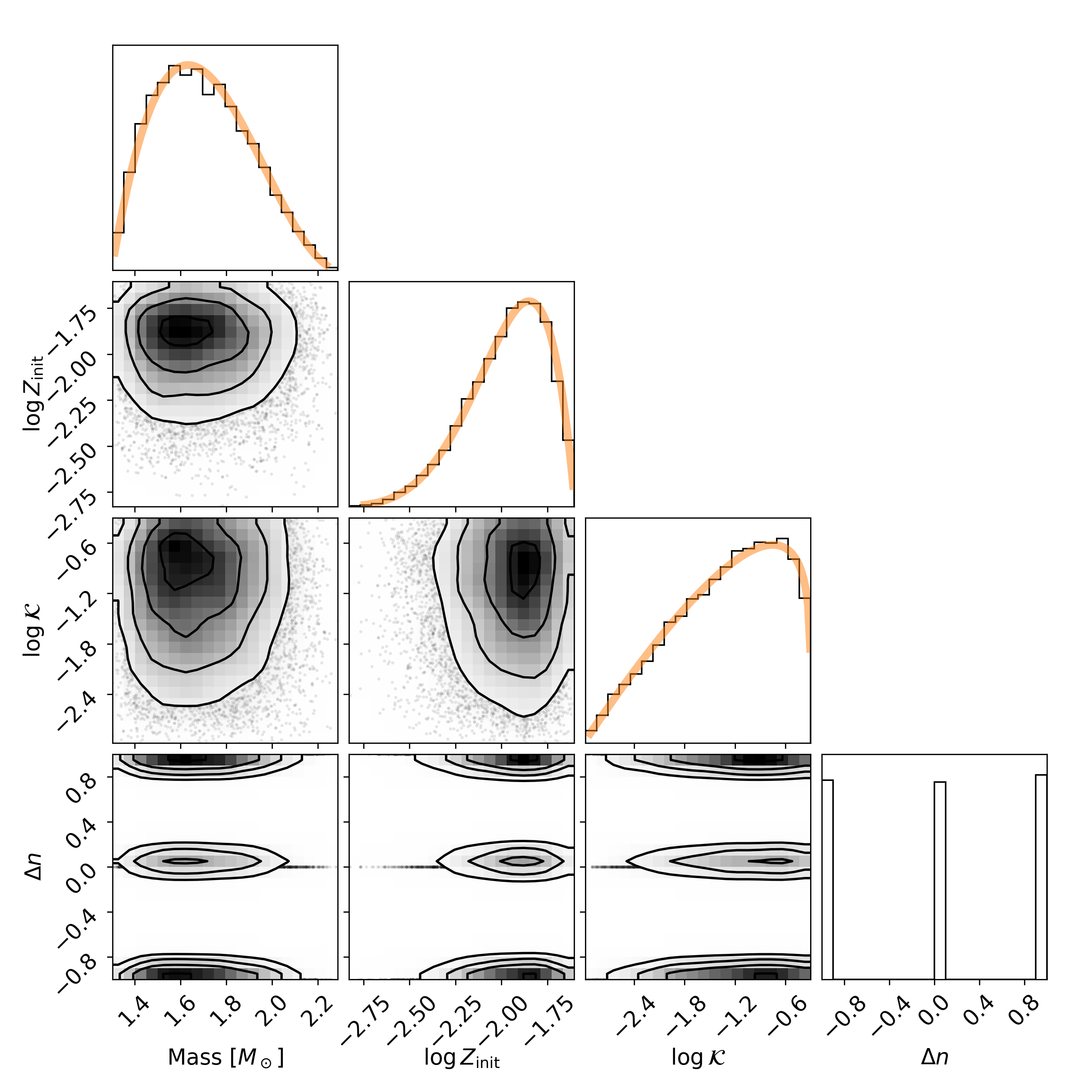}
    \includegraphics[width=0.43\linewidth, trim={0 0 20 0}, clip]{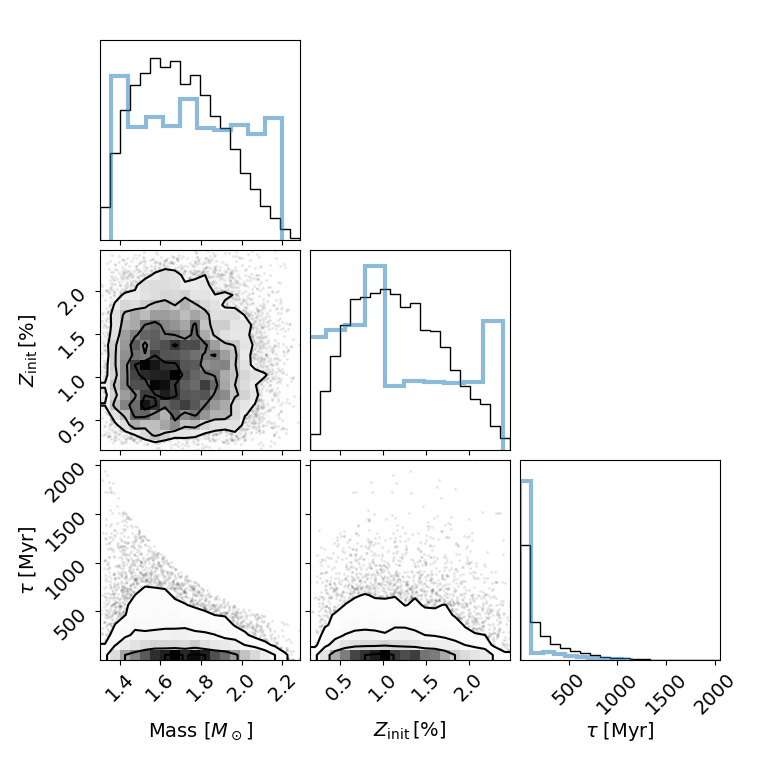}
    \caption{Left: Samples drawn from the one-dimensional prior distributions of the sampled parameters. The diagonal frames show the marginalized distributions (black) and the functions used to draw the samples (orange). The prior used for $\Delta n$ are $\delta$-functions at $\Delta n = -1, 0,\,\mathrm{and}\,1$. The off-diagonal frames show the two-dimensional distributions of the input parameters. Right: The samples from the left frames (black) transformed to the same units as the output from the stellar model grid (blue).}
    \label{fig:prior_samples}
\end{figure*}

Fig.\,\ref{fig:prior_samples} shows samples drawn from these prior density distributions, both in terms of the sampled variables and those transformed to $M$, \Zin\ and age. These priors are applied to the inference performed for all targets (see below). Additional priors may be applied on a target-by-target basis if, for example, the mass can be constrained by other sources such as orbiting companions, or limits can be placed on the metallicity by spectroscopy.

\begin{table}
	\centering
	\caption{Prior density functions used in Eq. \ref{eq:posterior}. The priors on $M$, $\log Z_{\mathrm{in}}$ and $\chron$ are given by $\beta^{a}_{b}$, where $a$ and $b$ are the shape parameters of the $\beta$-distributions, and the prior on $\Delta n$ is a series of $\delta$-functions at integer values. In all cases the arguments to the distribution functions denote lower and upper limits.}
	\label{tab:example_table}
	\begin{tabular}{cc} 
		\hline
		Parameter & Prior function\\
            \hline
            $M [M_{\odot}]$ & $\beta^{2}_{3}(1.3, 2.3)$ \\
            $\log Z_{\mathrm{in}}$ & $\beta^{6}_{2}(-3.1, -1.6)$ \\
            $\log \chron$ & $\beta^{2}_{1.2}(-3, -0.3)$ \\
            $\Delta n$ & $\in_R \{-1, 0, 1\}$\\
		\hline
	\end{tabular}
 \label{tab:priors}
\end{table}

For each of the samples drawn from the prior distributions, the \NN produces the following outputs: a set of mode frequencies, the effective temperature, and the luminosity. Given a set of outputs we then evaluated the likelihood of the observations by
\begin{equation}
    \log{\mathcal{L}\left(D|\theta\right)} = \log{\mathcal{L}(D_{\mathrm{S}}|\theta)} + \log{\mathcal{L}(D_{\mathrm{C}}|\theta)}.
\end{equation}
We separate the log-likelihood into the seismic variables, $D_\mathrm{S}$, and the classical (non-seismic) variables, $D_\mathrm{C}$. The contribution to the likelihood of the mode frequencies is given by
\begin{equation}
   \log{\mathcal{L}(D_{\mathrm{S}}|\theta)} = \sum_i \log \mathcal{N}\left( \nu^{\mathrm{obs}}_i, \sqrt{\sigma_{\nu^{\mathrm{obs}}_i}^2 + \sigma_{\nu^{\mathrm{NN}}_i}^2} \right),
\end{equation}
and that of the classical observables is given by
\begin{align}
   \log{\mathcal{L}(D_{\mathrm{C}}|\theta)} = &\log\mathcal{N}\left(\log{L}^{\mathrm{obs}}, \sqrt{\sigma_{L^{\mathrm{obs}}}^2 + \sigma_{L^{\mathrm{NN}}}^2}\right) \,+ \nonumber \\
   &\log\mathcal{N}\left(\Teff^{\mathrm{obs}}, \sqrt{\sigma_{T^{\mathrm{obs}}}^2 + \sigma_{T^{\mathrm{NN}}}^2}\right).
\end{align}
The width of the probability densities used in the inference is given by two terms that specify the observational uncertainty (superscript `obs'), and the noise due to the precision of the neural network's ability to emulate the model grid (superscript `NN'). Based on the spread of the residuals presented in Fig.\,\ref{fig:nn_res} and Fig.\,\ref{fig:unc_over_track}, this emulation uncertainty is approximately $4\times10^{-4}$ dex, which equates to a relative uncertainty of $\approx0.1$ per cent on the output parameters. This additional uncertainty was added in quadrature to the uncertainty of the observed mode frequencies. 

In the following we will use simulated frequencies corresponding to those obtained from $2$ sectors of data from the \textit{TESS} mission \citep{rickeretal2015}. We therefore adopted an uncertainty on the mode frequencies of $0.02\mathrm{d}^{-1}$, which is the frequency resolution of the resulting power spectra. The uncertainties on $\log L$ and $\log \Teff$ depend on the target in question, but for the simulations shown below these were fixed to $\sigma_L=0.05$ dex and $\sigma_T=200\mathrm{K}$. 

The \NN residuals also showed a bias of $\sim10^{-5}$ dex, which equates to an offset of $0.01$ per cent on each of the output parameters. This offset is small compared to the combination of the observed and \NN uncertainties, and so we did not consider it in the analysis. However, if either of these sources of uncertainty were decreased by, for example, improving the estimates of the observed mode frequencies, the importance of this bias would need to be re-evaluated. 

We performed the sampling using the nested sampling method from the \textsc{Dynesty} Python package \citep{Skilling2004, Speagle2020}. Nested sampling determines iso-likelihood contours in the input parameter space, which were iteratively redefined until samples were consistently drawn around the global likelihood maximum. In the \textsc{Dynesty} package, this process is terminated when the change in model log-evidence $\Delta \log \mathcal{E}$ is less than a predefined value chosen according to the \textsc{Dynesty} documentation\footnote{https://dynesty.readthedocs.io/en/stable/}. The method presented above is not restricted to using \textsc{Dynesty}, and other sampling methods may be used, such as \textsc{MultiNest} \citep{Buchner2014} or \textsc{EMCEE} \citep{Foreman-Mackey2013}.

\section{Results}
\label{sec:results}
\subsection{Simulated stars}
In order to test our methodology and validate the accuracy of the \NN emulator, we have performed tests based on 25 simulated stars in a `hare-and-hounds' exercise.  To produce these simulated stars, we proceeded as follows.  Values of stellar mass and initial metallicity were selected to lie in between values in the grid, but still within the defined parameter range of the grid.  
We calculated stellar models and pulsation frequencies using \textsc{MESA} and \textsc{GYRE}, using the same settings as for the grid.  
We selected ages from the newly calculated tracks, which then defined the truth values for mass, metallicity and age for our simulated stars and their associated `true' observables. We only selected a subset of the calculated modes, to better reflect typical observations of \dsct\ stars.  We selected modes at four consecutive radial orders for each degree, within the bounds of $n=1$--8. 

To simulate noisy observations, we added noise to the observable parameters of the simulated stars. These random offsets were drawn from a normal distribution, with mean of zero and a standard deviation of: $0.02$\,d$^{-1}$ for the mode frequencies, $200 \, \rm K$ for the effective temperatures, and $0.05 \, \rm dex$ for the log-luminosity. 

\subsubsection{Exemplar simulated stars}

\begin{figure*}
    \centering
    \includegraphics[width=.5\linewidth]{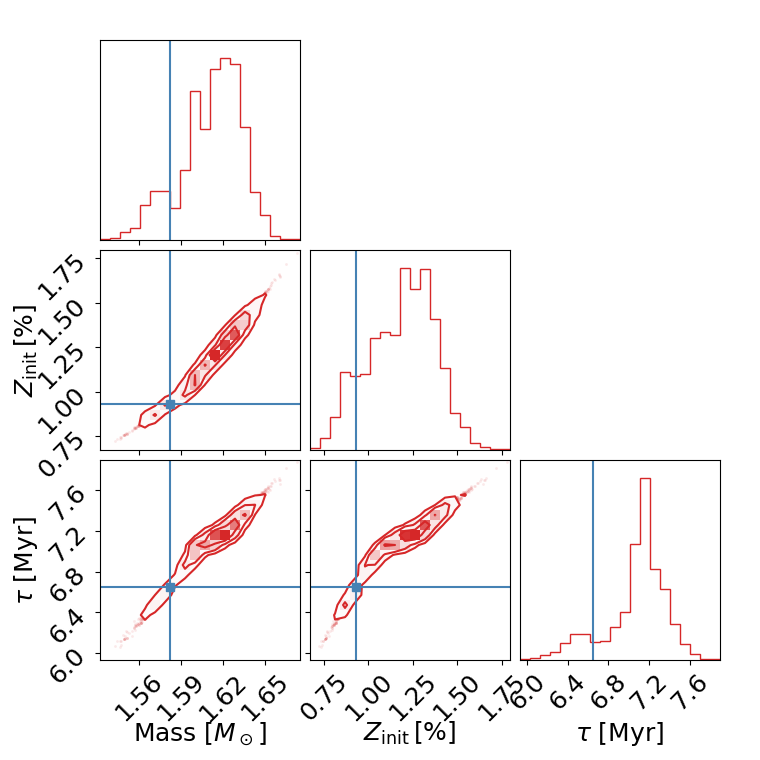}\hfill
    \includegraphics[width=.5\linewidth]{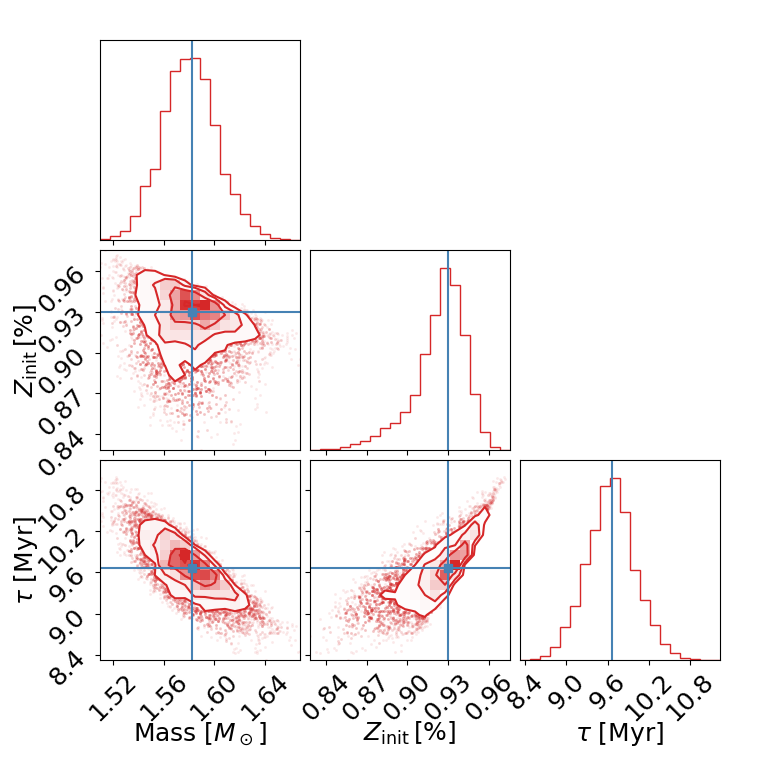}
    \caption{Samples drawn from the posterior distributions of exemplar simulated stars 5 (left) and 6 (right). For clarity we transform the initial metallicity to $\%$ and $\chron$ to stellar age $\tau$. The model input values used to generate the simulated star data are shown in blue.}
    \label{fig:exemplar_posterior_samples}
\end{figure*}

Fig.\,\ref{fig:exemplar_posterior_samples} shows the posterior probability estimates for simulated stars with some of the most promising results. The posterior samples for simulated star index 5 demonstrate our ability to quantify random uncertainty on our inferred properties from the posterior and shows that the true properties of this simulated star lie comfortably with the posterior distribution.  In this case our method is performing as required but it is worth noting that, even for this exemplar, the posterior still contains significant correlation between the parameters.

We see that the posterior distribution is not well described by a series of separable 1-D normal distributions.  Instead, there are strong covariances between inferred parameters, which are to be expected from stellar evolution theory.  However, the 1-D marginal posterior distributions show evidence of not being normally distributed and, in the case of the stellar age, even somewhat multi-modal. To examine the degree of accuracy and precision of the results we will study the summary statistics of the posterior distribution.  This does not capture all the detail that is of value, but is nonetheless useful as a test of our methods.

The posterior samples for simulated star index 6 show an ideal example of posterior probability estimation for a simulated star. Each model parameter posterior distribution is uni-modal, with peaks ($M = 1.58\,\Msun, Z_{\mathrm{init}} = 0.93\%, \tau = 9.65$\,Myr) comfortably centred on the truth values used to generate simulated star 6 ($M = 1.58\,\Msun, Z_{\mathrm{init}} = 0.93\%, \tau = 9.67$\,Myr).

\subsubsection{Results for 25 simulated stars}
\label{sec:simulated-stars-all}

While examining simulated stars individually is useful, it is hard to draw conclusions on the validity of our approach because we are looking at a single realisations of noise on the observables.  We now consider all 25 simulated stars, including the exemplars above (simulated star indices 5 and 6), to look at the statistics of our posterior probability distributions when compared to the truth values of the input properties. As part of our method, we fitted a parameter to account for our uncertainty in our assumption of the radial order label $n$.  In our tests on simulated stars, we recovered the correct radial order in all cases with no meaningful uncertainty on the posterior of the radial order labels.  

For each parameter of each star, we computed the difference between the truth value and the inferred value (the mean of the posterior samples for that parameter) divided by the uncertainty (the standard deviation of the posterior samples for that parameter).  If our inference is perfect, and our posterior distributions are well behaved, then this metric should be drawn from a normal distribution with zero mean and unit variance. However, multi modality, non-Gaussianity, and other pathological behaviour in the posterior can bias our metrics away from our assumed normal distribution.

\begin{figure*}
    \centering
    \includegraphics[width=.33\linewidth]{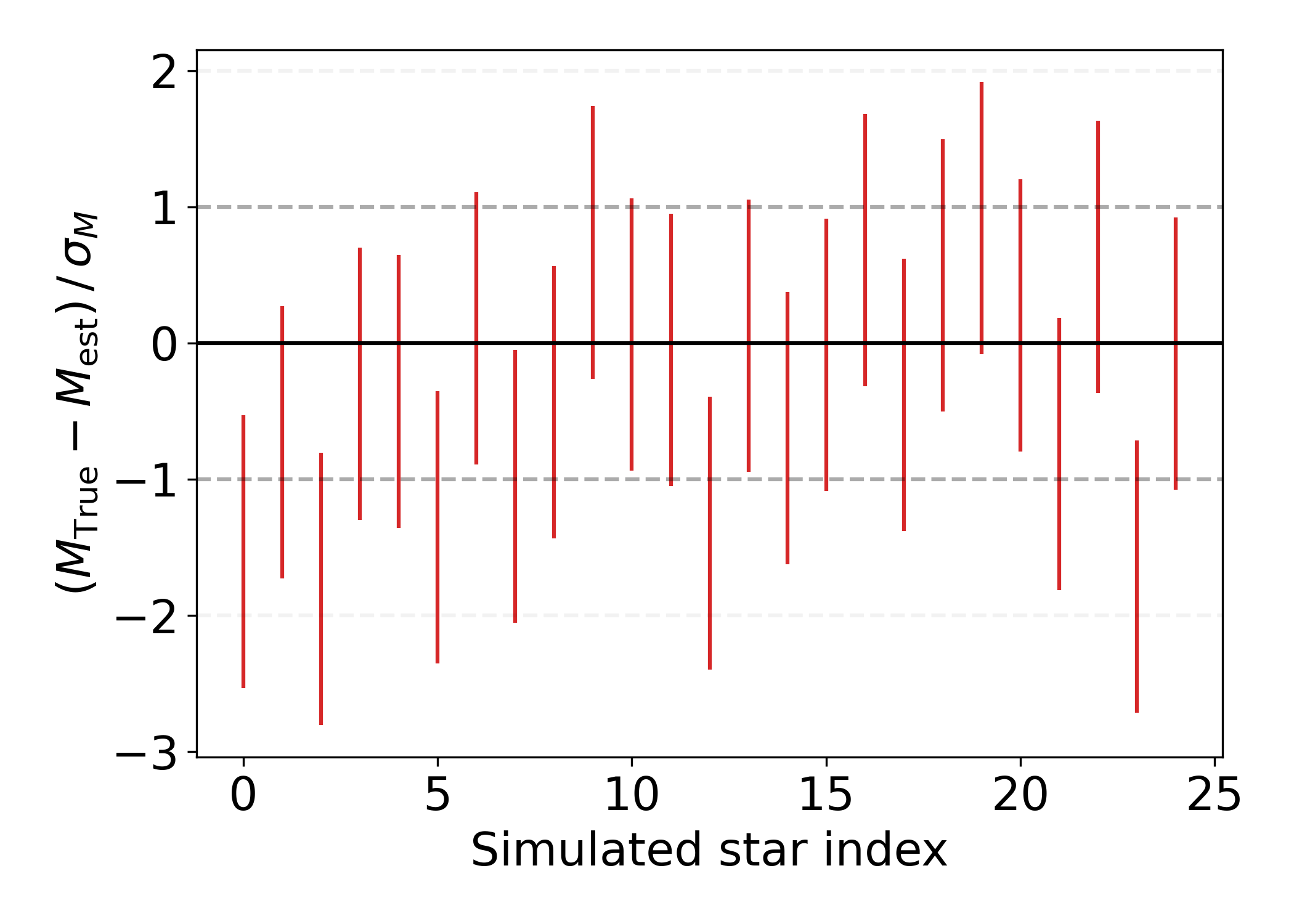}\hfill
    \includegraphics[width=.33\linewidth]{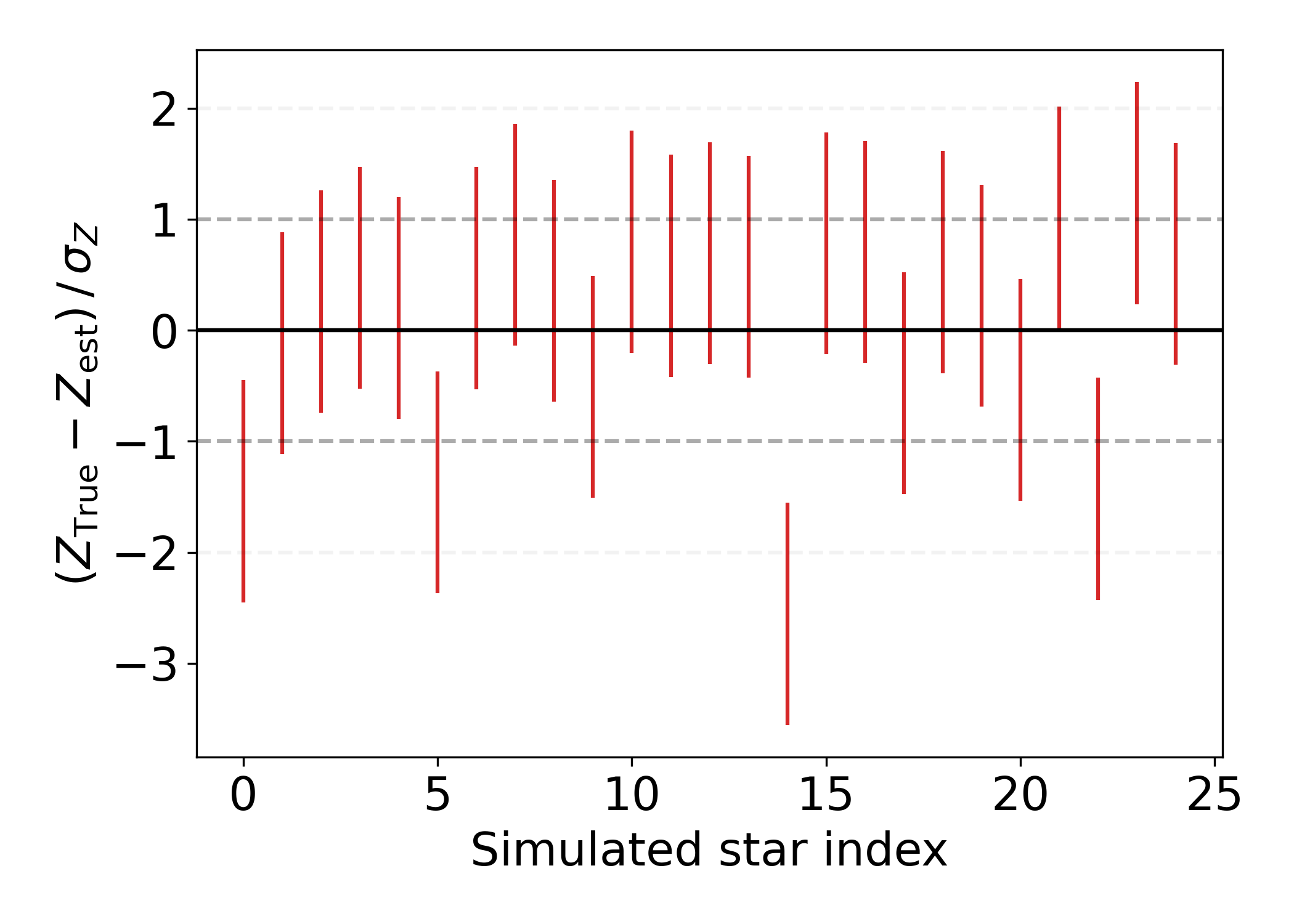}\hfill
    \includegraphics[width=.33\linewidth]{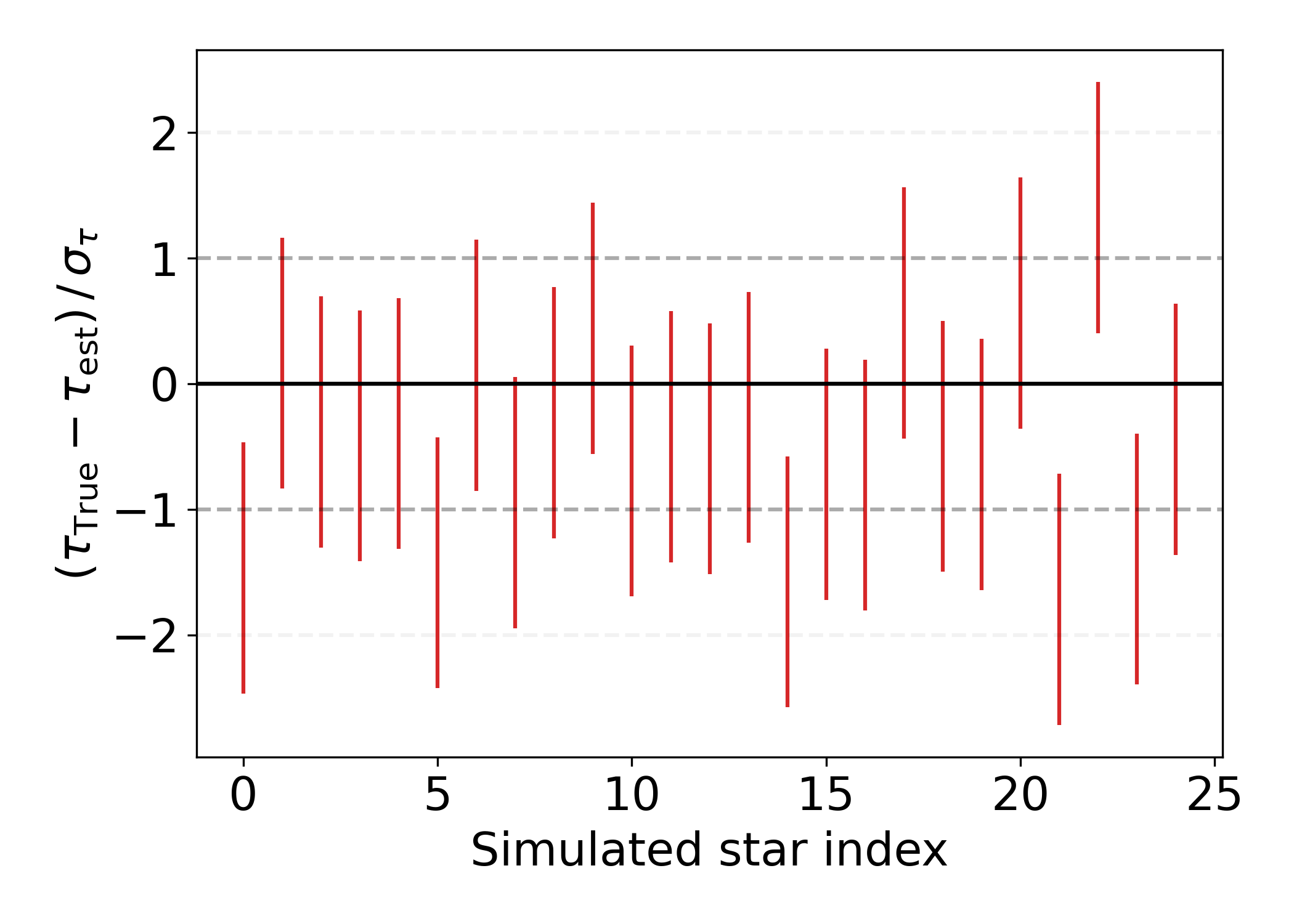}
    \caption{Difference between the inferred and model values of $M$, \Zin\ and age relative to the uncertainty of the inference, for a set of 25 simulated stars. The uncertainty is taken as the standard deviation of the marginalized posterior distributions of each of the parameters (see Sec.~\ref{sec:simulated-stars-all} for details). }
    \label{fig:hare_comparison}
\end{figure*}

Fig.\,\ref{fig:hare_comparison} shows the metric for each simulated star and each input property of the star.  It is clear that the majority of our simulated star results are consistent with the truth value given the uncertainty.  And broadly, the numbers of metrics at the 1 and 2 sigma levels are consistent with expectations. There are however some outliers or results which we will discuss.  The index for the most significant outlier in terms of metallicity is 14 and an interesting behaviour in the age distribution is observed in simulated star index 21 . 

\subsubsection{Further tests of simulated star 14}
\label{sec:simulated-star-14}

Simulated star 14 appears as an outlier in metallicity by $\sim2.5\sigma$. To examine this behaviour we have produced 10 more realisations of this simulated star.  That is, we have taken the same truth values as inputs, but redrawn the simulated noise on each observable parameter using the same noise distributions.

\begin{figure}
    \centering
    \includegraphics[width=\linewidth]{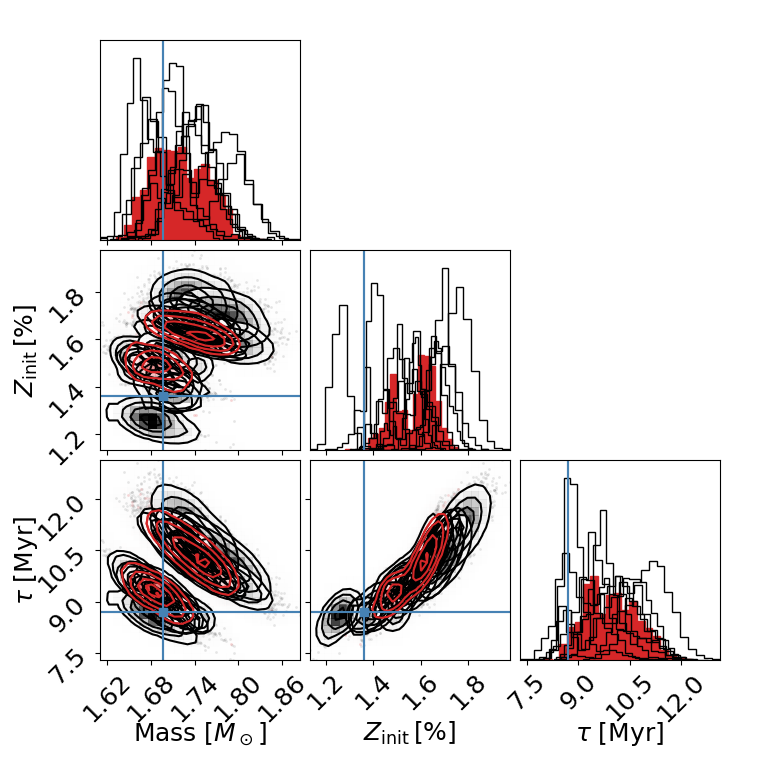}
    \caption{Samples drawn from the posterior distribution of simulated star 14 are shown in red.  Subsequent runs using the same truth values (M=1.696\,\Msun, \Zin=0.0136, $\tau$=8.76\,Myr) but different noise realisations are shown in black.  The truth values are shown in blue (see Sec.~\ref{sec:simulated-star-14} for details).}
    \label{fig:posterior_samples14_comp}
\end{figure}

Fig.\,\ref{fig:posterior_samples14_comp} shows the posterior samples for the original and subsequent runs for simulated star 14.  Firstly, it is clear that the posterior for \Zin is multi-modal and contains significant covariance.  Secondly, there is a bias of the posterior distributions away from the truth value in both metallicity and age that cannot be explained simply as a result of the realisation noise on the observables. 

We have checked for the possible origins of this bias.  We examined the prior probability distribution, but found it to be smooth and nearly flat over the region of the posterior.  We have performed multiple realisations of the noise and still observed this bias and therefore also exclude the noise or likelihood as the source of the bias. 

A possible source of error is in the neural network emulation producing differences in the predicted mode frequencies.  While these errors are typically small, of order $3\,\times\,10^{-4}$ dex, the noise from the neural network is not random noise that would be expected to reduce with more realisations of the observables.  Instead, the error is systematic and will produce a bias. The systematic error will always be present in emulation and this will lead to a bias, but it is the magnitude of this bias that is interesting.  For this simulated star, the bias is similar to the reported uncertainty, which is about $1.5 \, \rm Myr$. However, this error can be reduced by extending the training time of the \NN, or increasing the grid search density around the optimal \NN architecture. 

\subsubsection{Further tests of simulated star 21}
\label{sec:simulated-star-21}

\begin{figure}
    \centering
    \includegraphics[width=\linewidth]{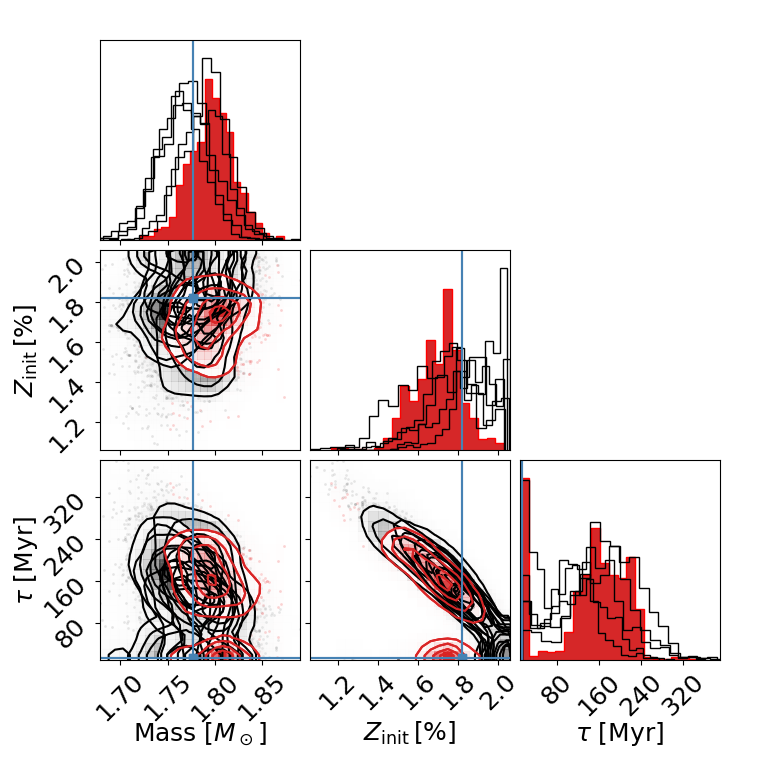}
    \caption{Samples drawn from the posterior distribution of simulated star 21 are shown in red.  Subsequent runs using the same truth values (M=1.777\,\Msun, \Zin=0.0182, $\tau$=12.97\,Myr) but different noise realisations are shown in black. The truth values are shown in blue (see Sec.~\ref{sec:simulated-star-21} for details).}
    \label{fig:posterior_samples21}
\end{figure}

Simulated star 21 shows an interesting behaviour in the age posterior. Fig.\,\ref{fig:posterior_samples21} shows the posteriors for the original simulated star 21 and for 10 more realisations, as we did for simulated star 14 above. No meaningful bias is observed in the posteriors for mass or metallicity, given the priors we apply.  

The true age of the simulated star was 12.96\,Myr, which corresponds to the pre-MS evolution stage.  The age posterior is clearly bimodal, with solutions around 10 Myr and 160 Myr.  This behaviour is observed in all the realisations, lending confidence that this is not a result of the noise being added.  In fact, this bi-modality is consistent with our understanding of the evolution of these stars and illustrates the difficulty of distinguishing the phase of the MS evolution where the track crosses its pre-MS evolution in the HR diagram.

\subsection{Application to HD\,99506}

\begin{figure}
    \centering
    \includegraphics[width=\linewidth]{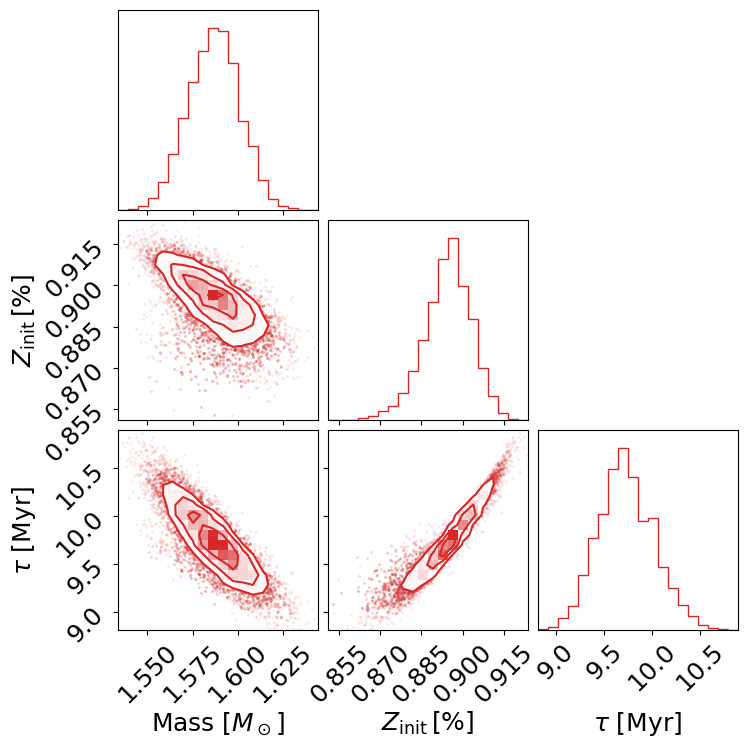}
    \caption{The posterior distributions in $M$, $Z_{\rm in}$ and age for HD\,99506, based on \textit{TESS} observations.}
    \label{fig:hd99506_corner}
\end{figure}

We applied our methods to HD\,99506, which is one of the high-frequency \dsct\ stars discussed by \citet{beddingetal2020}.
We used the following inputs: $T_{\rm eff} = 7970 \pm 250$\,K and $L$/L$_{\odot} = 7.58 \pm 0.37$ (taken from Table~1 of \citealt{beddingetal2020}), and the mode frequencies that we have measured and listed in Table\:\ref{tab:hd99506}. We chose only the mode frequencies that were obvious, leaving out tentatively identified modes such as the $n=4$ and $n=9$ radial modes, and the $n=1$ and $n=8$ dipole modes. The identified modes span two radial orders more than any of the simulations and so, despite the gaps at some orders, they provide tighter constraints. The resulting posteriors on $M$, $Z_{\rm in}$ and age are  unimodal and indicate a percent-level random uncertainty (Fig.\,\ref{fig:hd99506_corner}). The inferred age ($9.71\pm0.31$\,Myr) corresponds to the pre-MS phase, before the onset of pp-chain H-burning but after the temporary pre-MS CNO burning phase.

The calculation of well-sampled posteriors for uncertainty estimates is a marked improvement on what is possible using discrete grid points and $\chi^2$ minimisation (e.g. \citealt{kerretal2022a}), where an arbitrary threshold in $\chi^2$ needs to be adopted. It is especially useful that the posteriors are marginalized, given the aforementioned correlation in astrophysical parameters demonstrated with the simulated stars.

\begin{figure}
    \centering
    \includegraphics[width=\linewidth]{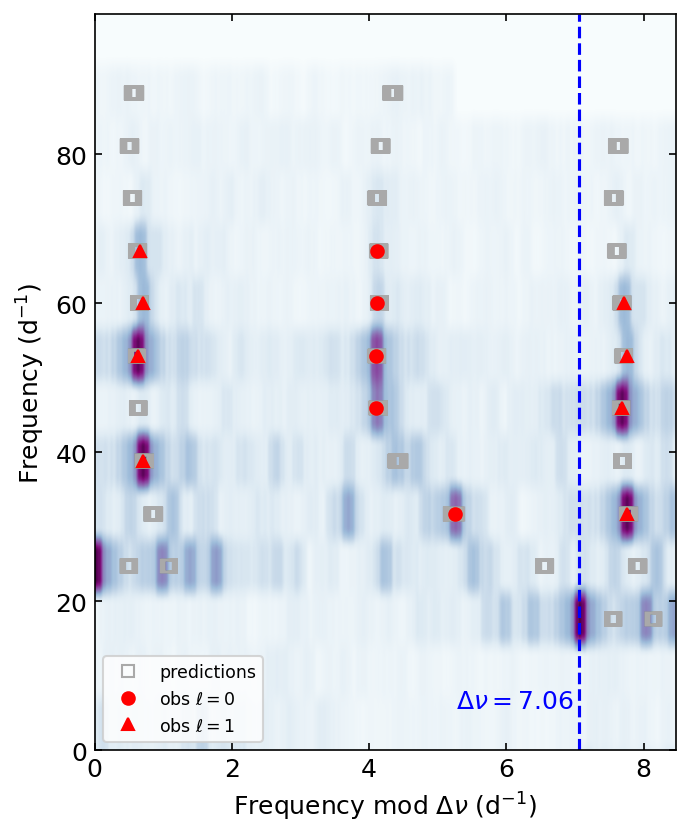}
    \caption{The posterior predicted frequencies overlaid on observed mode frequencies for HD\,99506. The greyscale is the observed amplitude spectrum smoothed by a Gaussian of width 4 times the frequency resolution. For each radial mode, multiple samples are taken from the retrieved model parameter posteriors and predictions for each sample are plotted.}
    \label{fig:hd99506_ech}
\end{figure}

\begin{table}
    \centering
        \caption{Identified modes for HD\,99506 used as inputs in the modelling.}
    \begin{tabular}{H c H H c c H}
    \toprule
$ID$&frequency&$a$&$p$&$n$&$\ell$&$m$\\
 & [d$^{-1}]$ & & & & & \\
 \midrule
$F5$&$33.48997$&$0.000848196824631597$&$0.441190534548$&$3$&$0$&$0$\\
$F6$&$46.46549$&$0.000734269706438689$&$0.8587626118250$&$5$&$0$&$0$\\
$F4$&$53.51968$&$0.00109673968124283$&$0.8101547896751$&$6$&$0$&$0$\\
$F21$&$60.60165$&$0.000132558002048196$&$0.87043014030$&$7$&$0$&$0$\\
$F31$&$67.65639$&$7.47281663326183e-05$&$0.66576768780$&$8$&$0$&$0$\\
$F2$&$35.99870$&$0.00179774746161778$&$0.41890141910223$&$3$&$1$&$0$\\
$F3$&$50.04504$&$0.00177988482166081$&$0.0010780190595$&$5$&$1$&$0$\\
$F11$&$57.18150$&$0.000316662125589266$&$0.58674908538$&$6$&$1$&$0$\\
$F12$&$64.19957$&$0.000317873066141502$&$0.29595053514$&$7$&$1$&$0$\\
\bottomrule
    \end{tabular}
    \label{tab:hd99506}
\end{table}

The neural network is also able to generate posterior predictions for each mode frequency, using the posterior samples as inputs. This can be useful for estimating the validity of uncertain mode identifications. In Fig.\,\ref{fig:hd99506_ech}, we see that the leftmost (lower frequency) of two close peaks at the $n=4$ radial mode is a good match and could perhaps be identified. The weak peak at $n=9$ would also have fitted well. Inclusion of these would have resulted in tighter posteriors. On the other hand, none of the missing dipole mode frequencies, nor the $n=1$ or 2 radial modes, would have been good additions. If we had supplied those modes as input, the posteriors would have broadened markedly.

\section{Conclusions}
We have presented a method for performing Bayesian inference on fundamental stellar properties of \dsct\ stars using a \NN. This method emulates the stellar model and oscillation codes, \textsc{MESA} and \textsc{GYRE}, by learning from a grid of models that encompasses the physical properties of stars in or near the \dsct\ instability strip. We used a nested sampling method to estimate the posterior distribution of the fundamental stellar properties, given a set of mode frequencies and classical observables. The resulting posterior distribution reflects the random observational uncertainty as well as the uncertainty of the \NN.  By providing samples from the posterior probability density, which might be multimodal, non-Gaussian, and strongly covariate, we formally quantified the statistical uncertainty in the fundamental stellar properties. This improves our ability to investigate the systematic uncertainty in the stellar models.

We used a test set that was initially unseen by the training algorithm to evaluate the performance of the trained \NN. We found that the \NN is capable of reaching an average frequency precision of $\approx3\times10^{-4}$ dex, with an offset of $\approx5\times10^{-5}$ dex. These performance metrics may improve if the network were re-trained with, for example, additional grid points or with the aim to reach a lower target loss. However, the flexibility of neural networks allows for the extension of the model grid to include additional variables, such as initial helium abundance or convective overshoot, by increasing the number of neurons in the network architecture \citep[see, e.g.,][]{Hendriks2019,Lyttle2021}. 

We applied our method to 25 simulated stars to quantify our accuracy and precision in the recovery of our input stellar properties. We showed the method to be capable of faithfully reproducing the true input parameters with the exception of simulated star 14. On investigation, we found a bias in the reported metallicity and age of this simulated star to be a result of the error in prediction by the neural network.  The bias is of order 1.5\,Myr in age for this star, but was confirmed to be systematic in nature. Further improvements in the accuracy of the neural network emulation would reduce the size of this effect. 

Finally, we have applied our method to observations of a real \dsct\ star, HD\,99506.  We found this star to be in the pre-MS stage of evolution and report a random uncertainty of only 3 per cent in age. 
Systematic uncertainty, such as that arising from missing or imperfect physics, has not been accounted for in this number but our methods pave the way for the quantification of the systematic uncertainty in future work. Similarly, the inclusion of additional physics such as rotation, and additional modes such as those of higher degree or different azimuthal order, would be trivial extensions to this framework in future.

\section*{Acknowledgements}
SJM was supported by the Australian Research Council (ARC) through Future Fellowship FT210100485.
MBN and GRD acknowledge support from the UK Space Agency.
TRB acknowledges support from Australian Research Council through Laureate Fellowship FL220100117.
OJS and AJL acknowledge the support of the Science and Technology Facilities Council.
This paper has received funding from the European Research Council (ERC) under the European Union’s Horizon 2020 research and innovation programme (CartographY GA. 804752).
This paper includes data collected by the \textit{TESS} mission. Funding for the \textit{TESS} mission is provided by the NASA's Science Mission Directorate.

\section*{Data Availability}
The data underlying this study will be supplied upon reasonable request to the authors.

\section*{Software}
Below we include additional software used in this work which has not explicitly been mentioned above.
\begin{itemize}
    \item Python \citet{python1995}
    \item matplotlib \citet{hunter2007}
    \item Numpy \citet{Harris2020}
    \item Scipy \citet{Virtanen2020}
    \item Pandas \citet{Reback2020}
    \item corner \citet{foreman-mackey2016}
    \item lightkurve \citet{lightkurvecollaboration2018}
    \item echelle \citet{hey&ball2020}
\end{itemize}
\bibliographystyle{mnras}
\bibliography{main}

\appendix

\bsp	
\label{lastpage}

\end{document}